\documentclass[aps,preprint,showpacs, showkeys]{revtex4}%
\usepackage{amsfonts}
\usepackage{amsmath}
\usepackage{amssymb}
\usepackage{graphicx}%
\providecommand{\U}[1]{\protect \rule{.1in}{.1in}}

\begin{document}
\title[ ]{ Comment on ``Families of Particles with Different Masses in $\mathcal{PT}%
$-Symmetric Quantum Field Theory'' }
\author{Abouzeid. M. Shalaby\footnote{E-mail:amshalab@ mans.edu.eg}}
\affiliation{Physics Department, Faculty of Science, Mansoura University, Egypt.}

\begin{abstract}
In a recent letter by Carl M. Bender and S. P. Klevansky an elementary field
theory is proposed where the authors claimed that it can account for the
existence of family of particles without any employment of group structures.
In this work, we relied on the fact that the Dyson-Schwinger equations used by
the authors stem from a variational principle and tried to reproduce their
results by a mere variational algorithm. We were able to reproduce exactly
their results by the employment of two different variational calculations of
the vacuum energy in the proposed theory. We showed that the two vacuum
solutions obtained by the authors of the commented letter correspond to two
non-degenerate vacuum energies except for the $d=0$ space-time dimensions
where the vacuum states are degenerate. Since any variational calculation of
the vacuum energy should be higher than the true (exact) vacuum energy, the
most accurate result will correspond to the vacuum that lowers the vacuum
energy. We showed that the vacuum with unbroken $Z_{2}$ symmetry has lower
energy than the vacuum with broken $Z_{2}$ symmetry. Accordingly, the theory
will prefer to live with the vacuum of unbroken $Z_{2}$ symmetry in which the
proposed theory is Hermitian. Thus the proposed theory neither has a preferred
non-Hermitian representation nor it describes a family of particles as claimed
by the authors of the commented work.

\end{abstract}
\maketitle

The discovery of possible physical applications of some non-Hermitian theories
may lead to a non-problematic description of natural events. The fruitfulness
of such trend varies from the discovery of unknown matter phases \cite{Novel}
to the possible solution of great puzzles like the Hierarchy puzzle
\cite{aboebt,ghost}. Very recently, the authors of Ref. \cite{bendfam}
introduced an idea by which they claimed that it can even account for the
description of different particle species without the resort to the usual
recipe of employing the group theory to the quantum field theory. In this
work, we show that their idea does not work for space-time dimensions greater
than zero. Even in the zero dimensional case, there exist only one
renormalized mass.

In Ref.\cite{bendfam}, Carl M. Bender and S. P. Klevansky showed that there
exist two vacuum solutions for the massless $\phi^{6}$ theory. The two
solutions obtained are characterized by two different masses. In their study,
they relied on the possible existence of many solutions to the Dyson-Schwinger
equations for different boundary conditions on the path of integration in
complex field space. First of all, they defined the renormalized masses at
different scales. The scale (or subtraction point) once chosen in a specific
calculation should be fixed for all other calculations. This criteria is known
in the literature by fixing the renormalization scheme \cite{collinbook}. In
fact, the formulae given by the authors for the renormalized mass describe
many masses ( not just two). This is because they defined the renormalized
mass at zero vacuum condensate which is described by a fixed scale while the
other mass defined at any vacuum condensate which means that this result will
give different masses at different scales. In general, a specific vacuum
condensate means a specific scale or subtraction point. Accordingly, defining
the renormalized mass for unspecified vacuum condensate is meaningless ( see
for instance the definition of the renormalized mass at some specific vacuum
condensate in Ref.\cite{Kaku}) . The second problem with the commented article
is that the Dyson-Schwinger equations stem from a variational principle and
although they might have many vacuum solutions, each vacuum will possess a
corresponding variational vacuum energy. However, a well known fact is that
any variational calculation of the ground state energy is higher than the true
(exact) vacuum energy. Accordingly, the most accurate result among the
possible vacuum solutions is the one that lowers the vacuum energy. In fact,
we used to have many vacuum solutions in quantum field theory which we call
them different phases of the theory. However, at some specific scale, the
theory will prefer one vacuum solution except at the critical point where
degenerate vacua may exist. This trend in understanding the existence of
different vacuum solutions in quantum field theory has been asserted
experimentally via the measurement of critical exponents for magnetic systems
(for instance) that lie in the same class of universality with a specific
quantum field theory ( like Ising model and $\phi^{4}$ theory)
\cite{physrep,zinjustin,guil,berzin}.

Rather than the above mentioned known facts about certain terminologies that
are clear in literature, in this work, to reinforce our reasoning in a
calculational manner, we follow a clear variational calculations that
reproduce the same results presented in Ref.\cite{bendfam}. However, the two
different variational calculations we use will give two different vacuum
energies. As we explained above, the theory will prefer to live with the
vacuum that lowers the vacuum energy. In that sense, the theory does have one
and only one acceptable vacuum solution and thus describe only one particle
and not a family of particles as claimed by the authors of Ref. \cite{bendfam}.

To start, consider the Hamiltonian density of the form;%
\begin{equation}
H=\frac{1}{2}\left(  \triangledown \phi \right)  ^{2}+\frac{1}{2}\pi^{2}%
+g\phi^{6}, \label{ham1}%
\end{equation}
where the field $\phi$ has the conjugated momentum field $\pi$ while $g$ is a
coupling constant.

Let us rewrite this form in terms of a variational mass $M$ as;%
\begin{align}
H  &  =\frac{1}{2}\left(  \triangledown \phi \right)  ^{2}+\frac{1}{2}\pi
^{2}+\frac{1}{2}M^{2}\phi^{2}+g\phi^{6}-\frac{1}{2}M^{2}\phi^{2},\nonumber \\
&  =H_{0}+H_{I}.
\end{align}
Here $H_{0}$ represents the free Hamiltonian $\frac{1}{2}\left(
\triangledown \phi \right)  ^{2}+\frac{1}{2}\pi^{2}+\frac{1}{2}M^{2}\phi^{2}$
while $H_{I}$ represents the interaction Hamiltonian of the form $g\phi
^{6}-\frac{1}{2}M^{2}\phi^{2}$. The vacuum energy of this theory is defined
as;%
\[
E_{0}\left(  M,g\right)  =\langle0\left \vert H\right \vert 0\rangle,
\]
which up to first order in the coupling has the form;%
\begin{align}
E_{0}\left(  M,g\right)   &  =-\frac{1}{2^{d+1}}\pi^{-\frac{d}{2}}M^{d}%
\Gamma \left(  -\frac{1}{2}d\right)  +15g\Delta^{3}-\frac{1}{2}M^{2}%
\allowbreak \Delta,\nonumber \\
\Delta &  =\frac{1}{\left(  4\pi \right)  ^{\frac{d}{2}}}\frac{\Gamma \left(
1-\frac{d}{2}\right)  }{\Gamma \left(  1\right)  }\left(  \frac{1}{M^{2}%
}\right)  ^{1-\frac{d}{2}}, \label{EO}%
\end{align}
where $d$ is the dimension of the space-time and $\Gamma$ is the gamma
function. In applying the variational condition of the form $\frac{\partial
E_{0}\left(  M,g\right)  }{\partial M}=0$, we get the result;
\begin{equation}
\left(  45h\Delta^{2}-\frac{1}{2}M^{2}\right)  \left(  -2^{1-d}M^{d-3}%
\Gamma \left(  2-\frac{1}{2}d\right)  \pi^{-\frac{1}{2}d}\right)  =0.
\label{MO}%
\end{equation}
This equation gives the result;%
\begin{equation}
M_{0}\left(  g\right)  =\exp \left(  \frac{1}{2}\frac{\ln \frac{1}{90g\left(
\Gamma \left(  1-\frac{1}{2}d\right)  \right)  ^{2}}+d\ln4\pi}{d-3}\right)  ,
\label{MOF}%
\end{equation}
where $M_{0}$ is the mass in the unbroken symmetry phase.With this form of the
mass, the vacuum energy will depend on the coupling $g$ as well as some unit
mass $\mu$ that may arise from logarithmic divergences coming from
singularities of the gamma function.

Now, let us follow another variational calculation for the vacuum energy. This
time we use two variational parameters, the effective field mass $M$ and the
vacuum condensate $B$. In this case, the Hamiltonian in Eq.(\ref{ham1}) takes
the form;%
\begin{equation}
H=\frac{1}{2}\left(  \triangledown \psi \right)  ^{2}+\frac{1}{2}\Pi^{2}%
+\frac{1}{2}M^{2}\psi^{2}+\left(  15gB^{4}-\frac{1}{2}M^{2}\right)  \psi
^{2}+20gB^{3}\allowbreak \psi^{3}+15gB^{2}\psi^{4}+6Bg\psi^{5}+g\psi^{6}%
+gB^{6},
\end{equation}
where we used the canonical transformation $\phi=\psi+B$ and $\pi=\Pi$. Here,
$B$ is a constant called the vacuum condensate and $\Pi=\overset{\cdot}{\psi}%
$. Moreover, we dropped out the linear term in the field $\psi$ since the
stability condition will kill it out order by order \cite{Peskin}. Up to first
order in the coupling $g$, the vacuum energy takes the form \cite{abostab};%
\begin{equation}
\allowbreak E_{B}\left(  M,B,g\right)  =gB^{6}+15\Delta gB^{4}+45gB^{2}%
\Delta^{2}-\frac{1}{2}\pi^{-\frac{1}{2}d}\Gamma \left(  -\frac{1}{2}d\right)
2^{-d}M^{d}-M^{2}\Delta-30g\Delta^{3}.\label{EB}%
\end{equation}
In applying the variational conditions $\frac{\partial E_{B}\left(
M,B,g\right)  }{\partial M}=0$ and $\frac{\partial E_{B}\left(  M,B,g\right)
}{\partial B}$, we get;%
\begin{align}
\frac{\partial E_{B}}{\partial M} &  =\frac{-2^{-1-d}\Gamma \left(  -\frac
{1}{2}d\right)  d\left(  -2+d\right)  M^{d-3}\pi^{-\frac{1}{2}d}}{2}\left(
30gB^{4}+180gB^{2}\Delta-M^{2}+90g\Delta^{2}\right)  =0,\nonumber \\
\frac{\partial E_{B}}{\partial M} &  =B\left(  6B^{4}+60\Delta gB^{2}%
+90g\Delta^{2}\right)  =0.
\end{align}
For $B=0$, we get exactly the same result in Eq.(\ref{MO}). On the other hand,
for $B\neq0$, we have the conditions;
\begin{align}
30gB^{4}+180gB^{2}\Delta-M^{2}+90g\Delta^{2} &  =0,\nonumber \\
\left(  6B^{4}+60\Delta gB^{2}+90g\Delta^{2}\right)   &  =0,\label{MB}%
\end{align}
where they coincide with the two conditions obtained in Ref.\cite{bendfam}. In
solving these two equations we get;%
\begin{align}
M_{B}\left(  g\right)   &  =\exp \left(  \frac{1}{2}\frac{\ln \left(  -\frac
{1}{30gd^{2}\left(  \Gamma \left(  -\frac{1}{2}d\right)  \right)  ^{2}\left(
-5+3e^{-d\ln4}4^{d}-\sqrt{10}\right)  }\right)  +d\ln4\pi}{-3+d}\right)
,\nonumber \\
B\left(  g\right)   &  =\left(  \frac{1}{60}2^{d}\frac{\pi^{\frac{1}{2}d}%
}{\Gamma \left(  -\frac{1}{2}d\right)  }\frac{M_{B}^{-d+2}}{dg}\left(
M_{B}^{2}+\frac{90}{4^{d}\pi^{d}}\left(  \Gamma \left(  -\frac{1}{2}d\right)
\right)  ^{2}M_{B}^{2d-4}d^{2}g\right)  \right)  ^{\frac{1}{2}}.\label{MBB}%
\end{align}
where we labeled the mass by the subscript $B$ to refer to the broken symmetry
($B\neq0$) case. Since the formulae in Eq.( \ref{MB}) are the same equations
obtained in Re.\cite{bendfam}, there is no surprise to see that our formulae
reproduce the same mass ratios since we have;
\begin{equation}
\frac{M_{B}}{M_{0}}=\exp \left(  -\frac{1}{2}\frac{2\ln2-\ln3+\ln \left(
2+\sqrt{10}\right)  }{-3+d}\right)  .
\end{equation}
For $d=0,.5,1,1.5,2$ and $2.5$, this equation gives $\frac{M_{B}}{M_{0}%
}=1.\, \allowbreak379\,2,$ $1.\, \allowbreak470\,8,1.\, \allowbreak
619\,7,1.\, \allowbreak902\,2,2.\, \allowbreak623\,6$ and $6.\, \allowbreak883$
respectively which are exactly the same values presented in Re. \cite{bendfam}%
, however, our calculations accounts for the vacuum energies as well. Using
the formula of $E_{0}$ in Eq.(\ref{EO}) with the corresponding mass given by
Eq.(\ref{MOF}) and the formula of $E_{B}$ from Eq.(\ref{EB}) with the mass and
vacuum condensate are given by Eq.(\ref{MBB}), one can get the result;
\begin{equation}
\frac{E_{B}}{E_{0}}=\exp \left(  -\frac{d}{2}\frac{2\ln2-\ln3+\ln \left(
2+\sqrt{10}\right)  }{-3+d}\right)  ,
\end{equation}
which gives $\frac{E_{B}}{E_{0}}=1,1.\, \allowbreak212\,8,1.\, \allowbreak
619\,7,2.\, \allowbreak623\,6,6.\, \allowbreak883$ and $124.\, \allowbreak29$ for
$d=0,.5,1,1.5,2$ and $2.5$ respectively. This means that the phase with
unbroken symmetry and that with broken symmetry are degenerate only at $d=0$.
For higher dimensions, the two vacua are non-degenerate and since we followed
a direct variational calculations, the most accurate result is that with lower
vacuum energy. In other words, the theory will prefer the vacuum with unbroken
symmetry ($B=0$) which means that the theory does describe one and only one
particle, not a family of particles as claimed by the authors of Ref.
\cite{bendfam}. Also, our result predict that the phase in which the theory is
non-Hermitian and $\mathcal{PT}$ -symmetric is not the preferred phase and
thus the theory is Hermitian in the Dirac sense. We need to assert that we
followed a clear variational calculations but we get exactly the same results
in the commented reference. The reason behind our choice for the variational
calculations is to show that the Dyson-Schwinger equations \ are in fact
equivalent to these variational calculations as expected since they stem from
variational ansatz.

To conclude, we have shown that the two vacuum solutions obtained in
Ref.\cite{bendfam} for the massless $\phi^{6}$ theory correspond to two
different variational calculations of the vacuum energy. Since any variational
calculation of the ground state energy is higher than the true vacuum energy,
the theory will prefer to live with the vacuum that lowers the vacuum energy.
We found that the vacuum with unbroken $Z_{2}$ symmetry is the preferred one
as it has the lowest energy among the available vacuum solutions. Since the
theory is Hermitian in the preferred phase and non-Hermitian in the other
available phase, we conclude that the theory under consideration does not have
an acceptable non-Hermitian representation. Moreover, since the vacua are
non-degenerate except for the $d=0$ case, the theory has only one acceptable
vacuum and thus it describes one particle only, not a family of particles as
claimed by the authors of Ref. \cite{bendfam}.

\end{document}